\documentclass[aip,pop,reprint,longbibliography,showkeys]{revtex4-1}

\usepackage{graphicx}
\usepackage{dcolumn}
\usepackage{bm}

\usepackage{mathptmx}      

\usepackage[version=3]{mhchem}
\usepackage{color}
\usepackage{amsmath}
\usepackage{booktabs,caption}
\usepackage[flushleft]{threeparttable}
\usepackage{multirow} 
\usepackage{nicefrac}
\usepackage{textcomp}

\newcommand{\DB}[1]{{\color{black}#1}}

\newcommand{\kscope}{Ka\-lli\-ro\-scope}

\begin{document}

\preprint{AIP/POF18-AR-01206}

\title{Rheoscopic Fluids \DB{in} a Post-\kscope\ World}

\author{Daniel Borrero-Echeverry}
\email{dborrero@willamette.edu}
\affiliation{Department of Physics, Willamette University, 900 State Street, Salem, OR 97301, USA}

\author{Christopher J. Crowley}%
\affiliation{ Center for Nonlinear Science and School of Physics, Georgia Institute of Technology, \\ \hspace{0.1in}837 State Street, Atlanta, GA 30332, USA}%

\author{Tyler P. Riddick}
\affiliation{Department of Physics, Willamette University, 900 State Street, Salem, OR 97301, USA}

\date{\today}

\begin{abstract}
In rheoscopic flow visualization, the working fluid is seeded with small reflective flakes that align preferentially in the flow due to their anisotropy. This leads to directed light scattering, which can be exploited to distinguish qualitatively different regions of the flow. For the past four decades, the gold standard in rheoscopic flow visualization has been \kscope, a commercial product consisting of crystalline guanine particles. Recently, however, worldwide production of crystalline guanine has dropped precipitously, leading the \kscope\ Corporation to halt production in 2014. Here, we present a short survey of alternative rheoscopic flow visualization techniques and introduce an inexpensive rheoscopic fluid based on stearic acid crystals extracted from shaving cream, which has a performance similar to, and in certain respects superior to, \kscope.
\end{abstract}

\maketitle

\section{Introduction}
Developments in particle-based velocimetry and camera technology have allowed researchers to quantitatively probe the structure of complex flows in ways that were un\-i\-ma\-gin\-able a mere decade ago.\cite{Westerweel2013,Raffel2018} Despite these advances, flow visualization remains a central tool in the experimentalist's arsenal,\cite{Merzkrich1987,Smits2012} allowing researchers to highlight flow features that are cumbersome to extract from quantitative flow measurements. \DB{Flow visualization also plays a vital role in informing the design of experimental systems by providing immediate visual feedback of flow conditions, thereby allowing researchers to configure their quantitative flow probes optimally.} 

Rheoscopic fluids are a common technique for visualizing spatially-extended flow structures. In rheoscopic flow visualization, the working fluid is see\-ded with small, anisotropic, reflective particles. \DB{Due to their an\-i\-so\-tro\-py, the particles become preferentially aligned in the flow. Therefore, when they scatter light, they highlight qualitatively different regions of the flow.\cite{Weidman1989}} Although it is commonly stated in the literature that particles align with the shear planes in the flow, this is only technically true for pure shear flow. \DB{Starting with the pioneering work of Jeffery,\cite{Jeffery1922} researchers have shown that particle motions are much more complicated,\cite{Savas1985,Gauthier1998,Goto2011,Shi2014,Byron2015} which makes interpreting local flow conditions from rheoscopic flow visualizations difficult.} Despite this, rheoscopic fluids have played an important role in the development of many areas of fluid mechanics. Here, we provide a historical \DB{survey} of rheoscopic flow visualization techniques and introduce a new rheoscopic fluid based on stearic acid crystals \DB{extracted from shaving cream}.

\section{A Short Survey of Rheoscopic Flow Visualization Methods}
\label{sec:Background}
The use of rheoscopic fluids dates back at least to J.C. Wilcke, who in 1785 used burnt lime particles to visualize vortex breakdown in a stirred tank.\cite{Weidman1989} The technique was later employed by B\'{e}nard, who used graphite and aluminum flakes to visualize convection patterns in heated fluid layers.\cite{Wesfreid2006a} At the turn of the 20\textsuperscript{th} century, Prandtl used mica flakes suspended on the surface of a water tunnel to visualize the flow around solid bodies.\cite{Raffel2007} \DB{In 1956}, Schultz-Grunow and Hein used a suspension of aluminum flakes extracted from hammer paint to visualize the wavy vortex regime of Taylor-Couette flow,\cite{SchultzGrunow1956} a technique that was later famously employed by Coles.\cite{Coles1965} Variations of the aluminum flake technique remained the dominant rheoscopic flow visualization method up to the late 1970s.\cite{Weidman1989}

\begin{figure}[b]
\def\svgwidth{3.37in}
\begingroup%
  \makeatletter%
  \providecommand\color[2][]{%
    \errmessage{(Inkscape) Color is used for the text in Inkscape, but the package 'color.sty' is not loaded}%
    \renewcommand\color[2][]{}%
  }%
  \providecommand\transparent[1]{%
    \errmessage{(Inkscape) Transparency is used (non-zero) for the text in Inkscape, but the package 'transparent.sty' is not loaded}%
    \renewcommand\transparent[1]{}%
  }%
  \providecommand\rotatebox[2]{#2}%
  \newcommand*\fsize{\dimexpr\f@size pt\relax}%
  \newcommand*\lineheight[1]{\fontsize{\fsize}{#1\fsize}\selectfont}%
  \ifx\svgwidth\undefined%
    \setlength{\unitlength}{236.92468071bp}%
    \ifx\svgscale\undefined%
      \relax%
    \else%
      \setlength{\unitlength}{\unitlength * \real{\svgscale}}%
    \fi%
  \else%
    \setlength{\unitlength}{\svgwidth}%
  \fi%
  \global\let\svgwidth\undefined%
  \global\let\svgscale\undefined%
  \makeatother%
  \begin{picture}(1,0.67216129)%
    \lineheight{1}%
    \setlength\tabcolsep{0pt}%
    \put(0,0){\includegraphics[width=\unitlength,page=1]{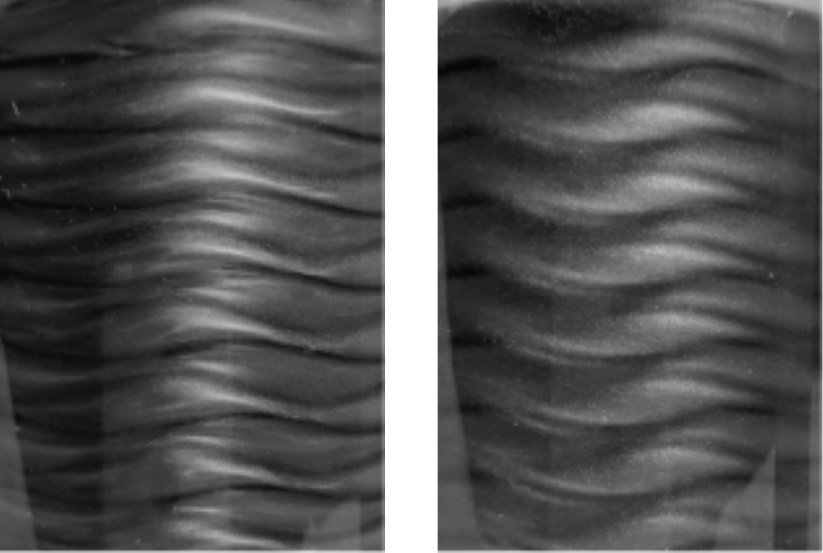}}%
    \put(0.01838592,0.0324674){\color[rgb]{1,1,1}\makebox(0,0)[lt]{\lineheight{1.25}\smash{\begin{tabular}[t]{l}(a)\end{tabular}}}}%
    \put(0.54988311,0.0324674){\color[rgb]{1,1,1}\makebox(0,0)[lt]{\lineheight{1.25}\smash{\begin{tabular}[t]{l}(b)\end{tabular}}}}%
    \put(0,0){\includegraphics[width=\unitlength,page=2]{Fig1.pdf}}%
    \put(0.87711793,0.0324674){\color[rgb]{1,1,1}\makebox(0,0)[lt]{\lineheight{1.25}\smash{\begin{tabular}[t]{l}7 mm\end{tabular}}}}%
    \put(0,0){\includegraphics[width=\unitlength,page=3]{Fig1.pdf}}%
    \put(0.34797113,0.0324674){\color[rgb]{1,1,1}\makebox(0,0)[lt]{\lineheight{1.25}\smash{\begin{tabular}[t]{l}7 mm\end{tabular}}}}%
  \end{picture}%
\endgroup%
\caption{\label{fig:KvsB}\DB{Rheoscopic flow visualization of wavy vortices in Taylor-Couette flow at a Reynolds number of $Re\sim 750$ using (a) \kscope\ and (b) rheoscopic fluid based on stearic acid crystals.}}
\end{figure}

In the mid-1960s, the artist Paul Matisse invented \kscope, a rheoscopic fluid based on suspensions of crystalline guanine, which he used to create ``kinetic sculptures.''\cite{Matisse1969} Although researchers used \kscope\ as early as 1970,\cite{Krishnamurti1970} initially the technique was not widely adopted.\cite{Borrero2014Thesis} However, in the early 1980s flow instabilities became popular testbeds for ideas emerging from the nascent field of nonlinear dynamics and \kscope\ quickly became a workhorse in the study of the spatiotemporal evolution of flow structures.  

The key to \kscope 's success is its crystalline guanine base. Crystalline guanine is a shiny substance extracted from fish scales.\cite{Pfaff1999} It is a particularly useful seeding material for rheoscopic fluids because it has a relatively low density (1.62 g/cm$^3$) and small particle size (6 $\times$ 30 $\times$ 0.07 $\mu$m$^3$). These properties cause guanine crystals to settle very slowly compared to other seeding materials. Guanine also has a high index of refraction (1.85), which makes it an effective scatterer, and leads to high-contrast \DB{flow visualization,\cite{Matisse1984} as shown in Fig. \ref{fig:KvsB}(a)}.  

Even though \kscope\ has been an important tool in the fluid dynamics community for over three decades, traditionally the main consumer of guanine pigments has been the cosmetics industry, which used them to produce pearlescence in a variety of consumer products. Guanine extraction, however, is a delicate and inefficient process requiring 50 tons of fish to be processed to produce 1 kg of guanine.\cite{Crombie1997} The resulting high cost of guanine pigments has driven the cosmetics industry to look for more inexpensive alternatives,\cite{Crombie1997,Pfaff1999} eventually leading to their adoption of pearlescent pigments based on mica flakes coated with various metal oxides (e.g., Merck \& Co.'s Iriodin family of pigments) or pearlizing agents based on glycol stearates.\cite{Crombie1997} As a result commercial production of guanine has declined, making guanine-based pigments such as Mearlmaid AA increasingly difficult to obtain. In 2014, the \kscope\ Corporation, citing the scarcity of raw materials, halted the production of \kscope\cite{Kscope2017} and left researchers scrambling for alternatives.

Aside from \kscope\ and the classical aluminum powder technique, other options for rheoscopic flow visualization include suspensions of commercial pigments based on mica flakes coated with metal oxides. While inexpensive, these pigments are formulated for use in the manufacture of paint and plastics, where the high viscosity of the fluid phase prevents their sedimentation. In fluid mechanics research, where less viscous working fluids are commonly used, mica-based pigments sediment relatively quickly. This sedimentation accentuates problems that emerge when using rheoscopic flow visualization to study flows with poor mixing properties or over long time scales.\cite{Dominguez-Lerma1985} More recently, biological materials such as tobacco mosaic virus\cite{Hu2009} and bacteria\cite{Marcos2011} have been employed. While the sedimentation rate for these materials is very low due to the small size of the particles, they are not commercially available and require significant infrastructure and expertise to grow in-house. \DB{A simple rheoscopic fluid based on stearic acid crystals extracted from shaving cream can circumvent these issues and, as can be seen in Fig. \ref{fig:KvsB}, has a performance similar to \kscope.}

\DB{Despite being used for well over a century, there is currently no consensus on what flow information rheoscopic flow visualization actually captures.} Some attempts have been made to establish correlations between images obtained using rheoscopic flow visualization and velocimetry data, but these are limited to specific geometries and flow regimes.\cite{Abcha2008} While it may be possible that no general interpretation of rheoscopic flow visualization images exists, recent progress has been made in visualizing data from computer simulations using ``virtual rheoscopic fluids.'' Here, numerically computed flow fields are used to determine the orientation that rheoscopic seeding particles would take in the flow. This information is then used to render visualizations that can be compared directly with experimental flow visualization images.\cite{Barth2007,Hecht2010,Goto2011}

\section{Preparing Rheoscopic Fluid From Shaving Cream}
\label{sec:Procedure}

\begin{figure}
\includegraphics[width=3.37in]{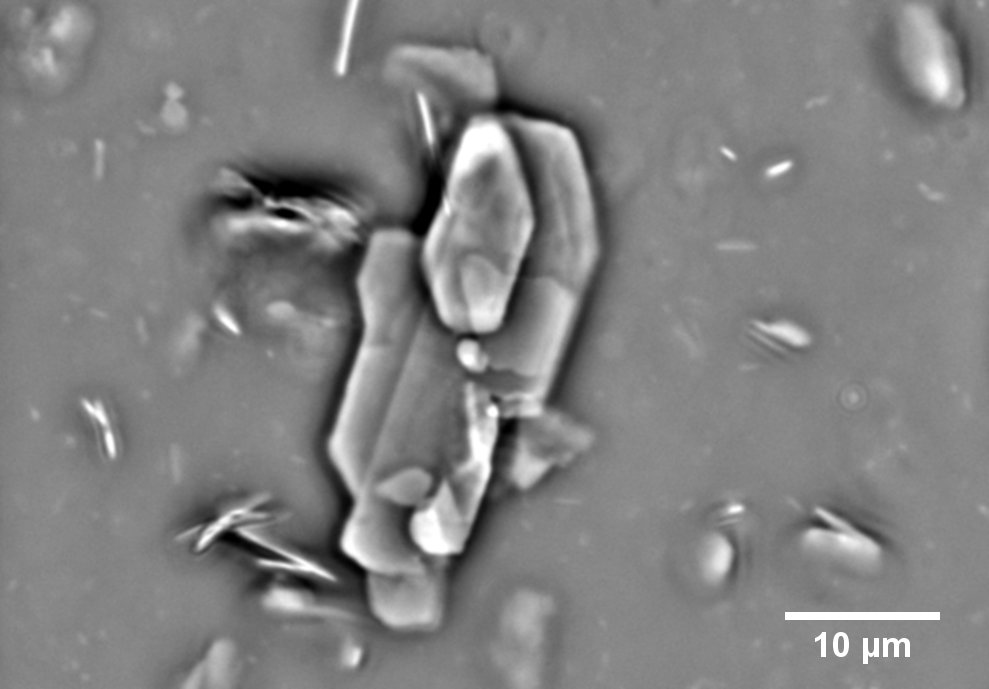}
\caption{\label{fig:CrystalPic}A phase-contrast micrograph of the rheoscopic fluid shows plate-shaped stearic acid crystals with a large distribution of sizes. \DB{The thinness of the crystals is apparent from edge-on crystals, which appear as white lines near the top and on the left side of the image.} The grayscale for the image has been inverted and the image contrast adjusted to highlight the crystal shapes.}
\end{figure}

\DB{We now discuss how to prepare a rheoscopic fluid using stearic acid crystals. The key ingredient in any rheoscopic fluid is its anisotropic seeding particles. It is possible to create stearic acid platelets by melting pure stearic and allowing it to crystallize from an aqueous solution.\cite{Zhu2007}} However, this process is tedious and requires precise control of the temperature during the crystallization process, as well as a careful selection of complementary surfactants to prevent the crystals from clumping. Luckily, the cosmetics industry has already solved these problems and stearic acid platelets, such as those shown in Fig. \ref{fig:CrystalPic}, can be easily ex\-trac\-ted from common shaving cream. This makes our rheoscopic fluid both inexpensive and easily accessible to scientists in developing countries, high-school teachers, and researchers at small institutions. 

\begin{figure*}
\includegraphics[width=6.67in]{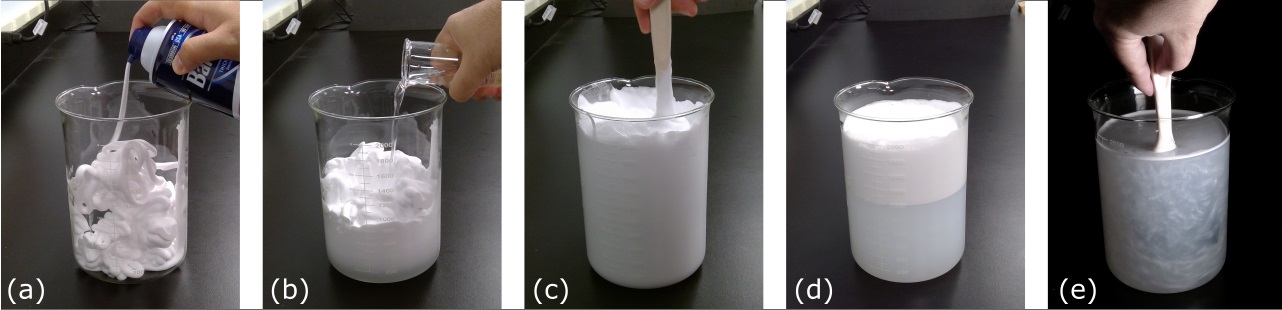}

\caption{\label{fig:Procedure}\DB{Preparation of rheoscopic fluid from shaving cream: (a) Shaving cream is added to a beaker. (b) Water is added at a ratio of 20:1 by mass. (c) The water and shaving cream are thoroughly mixed resulting in a thick, foamy liquid. (d) After a few hours, the mixture separates into a white foam and a milky, gray liquid. (e) After the foam is removed, the remaining liquid can be diluted and used for rheoscopic flow visualization.}}
\end{figure*}

\begin{figure}
\def\svgwidth{3.37in}
\begingroup%
  \makeatletter%
  \providecommand\color[2][]{%
    \errmessage{(Inkscape) Color is used for the text in Inkscape, but the package 'color.sty' is not loaded}%
    \renewcommand\color[2][]{}%
  }%
  \providecommand\transparent[1]{%
    \errmessage{(Inkscape) Transparency is used (non-zero) for the text in Inkscape, but the package 'transparent.sty' is not loaded}%
    \renewcommand\transparent[1]{}%
  }%
  \providecommand\rotatebox[2]{#2}%
  \newcommand*\fsize{\dimexpr\f@size pt\relax}%
  \newcommand*\lineheight[1]{\fontsize{\fsize}{#1\fsize}\selectfont}%
  \ifx\svgwidth\undefined%
    \setlength{\unitlength}{821.59715067bp}%
    \ifx\svgscale\undefined%
      \relax%
    \else%
      \setlength{\unitlength}{\unitlength * \real{\svgscale}}%
    \fi%
  \else%
    \setlength{\unitlength}{\svgwidth}%
  \fi%
  \global\let\svgwidth\undefined%
  \global\let\svgscale\undefined%
  \makeatother%
  \begin{picture}(1,1.08808302)%
    \lineheight{1}%
    \setlength\tabcolsep{0pt}%
    \put(0,0){\includegraphics[width=\unitlength,page=1]{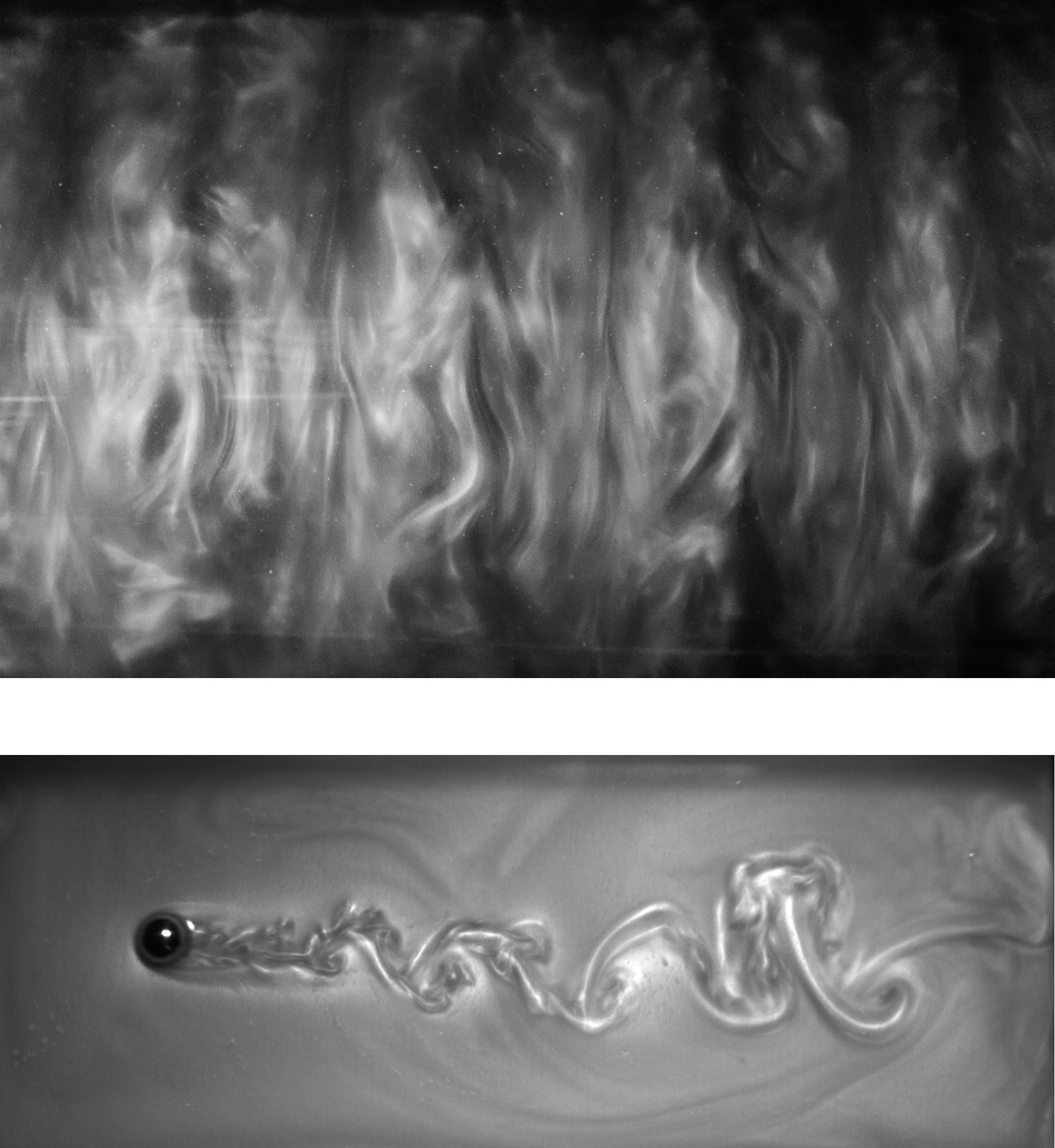}}%
    \put(0.02010209,0.47658264){\color[rgb]{1,1,1}\makebox(0,0)[lt]{\lineheight{1.25}\smash{\begin{tabular}[t]{l}(a)\end{tabular}}}}%
    \put(0.02010209,0.02606502){\color[rgb]{1,1,1}\makebox(0,0)[lt]{\lineheight{1.25}\smash{\begin{tabular}[t]{l}(b)\end{tabular}}}}%
    \put(0.84847813,0.02371741){\color[rgb]{1,1,1}\makebox(0,0)[lt]{\lineheight{1.25}\smash{\begin{tabular}[t]{l}25 mm\end{tabular}}}}%
    \put(0,0){\includegraphics[width=\unitlength,page=2]{Fig4.pdf}}%
    \put(0.83100393,0.47423529){\color[rgb]{1,1,1}\makebox(0,0)[lt]{\lineheight{1.25}\smash{\begin{tabular}[t]{l}15 mm\end{tabular}}}}%
    \put(0,0){\includegraphics[width=\unitlength,page=3]{Fig4.pdf}}%
  \end{picture}%
\endgroup%

\caption{\label{fig:FlowViz}Examples of flow visualization using stearic a\-cid\--based rheoscopic fluid: \DB{(a) Turbulent structures in a Taylor-Couette flow at $Re \sim 1600$. (b) Wake behind a steel ball rolling through a thin layer of fluid at $Re \sim 2000$.}}
\end{figure}

\DB{The extraction of stearic acid crystals, outlined in Fig. \ref{fig:Procedure}, begins by thoroughly mixing water and shaving cream at a ratio of 20:1 by mass.\footnote{\DB{This is approximately 1:1 by volume but measuring the volume of shaving cream is difficult, so we provide a more precise formulation in terms mass.}}} This ratio is not critical, but increasing the concentration of shaving cream beyond this point can lead to clumping of the stearic acid crystals. After letting the mixture settle for a few hours, it separates into a milky, gray liquid covered by white foam. Because the density of stearic acid is close to that of water, the stearic acid crystals remain suspended in the liquid, which can then be siphoned out from beneath the foam. Care should be taken not to siphon the heavy solids and larger clumps that settle at the bottom of the container. The resulting fluid is fairly concentrated and can be diluted as needed. It is possible to dissolve the leftover foam a second time to obtain additional rheoscopic fluid, but the pearlescence of these secondary batches is not as strong. \DB{Fig. \ref{fig:FlowViz} shows some examples of flow visualization using stearic acid-based rheoscopic fluid.}

We tested different brands of shaving cream and found that Barbasol Original Thick \& Rich Shaving Cream\texttrademark\ not only provides the best results, but conveniently only costs about US \$1-2 per can in bulk. Each 10 oz. can of Barbasol yields several gallons of useful rheoscopic fluid. Researchers in regions where Barbasol is not available should look for inexpensive shaving cream whose ingredients are water, stearic acid, triethanolamine, laureth-23, sodium lauryl sulfate, and fragrance. More complex formulations such as shaving gels or mentholated shaving cream do not work as well. 

\DB{The stearic acid crystals that form the basis for our rheoscopic fluid, appear to be an incidental byproduct of the manufacturing process of shaving cream. Unlike glycol stearates and mica flakes, which are deliberately added to shampoos and soaps as pearlizing agents,\cite{Crombie1997} stearic acid is added to shaving cream as an emollient and thickening agent.\cite{Krishan2017} However, at various points in the manufacturing process the shaving cream is heated to elevated temperatures (e.g., when testing the integrity of the cans before shipping).\cite{EAF2009} In the presence of triethanolamine and other surfactants, this is sufficient to dissolve the stearic acid in the water that makes up majority of the shaving cream. This sets up conditions similar to the experiments of Zhu et al., who showed that as an aqueous solution of stearic acid and triethanolamine cools, flat stearic acid crystals form through a nucleation and ripening process.\cite{Zhu2007}  In order to confirm that the solids in our fluid are made of stearic acid, we compared their nuclear magnetic resonance spectrum to publicly available reference spectra and found them in good agreement. Interested readers will find the physicochemical and crystallographic properties of these crystals in the review paper by Sato.\cite{Sato1987}} 

\section{Properties of Rheoscopic Fluids Based on Stearic Acid Crystals}
\label{sec:Discussion}

\DB{As seen in Fig. \ref{fig:CrystalPic}, the raw concentrate contains particles in a range of sizes, ranging from usefully large stearic acid platelets with extended dimensions in the 5-50 micron range and thicknesses of about 0.3 microns to much smaller, more isotropic solids with dimensions in 0.1-1 micron range. These small solids (visible as small specks in Fig. \ref{fig:CrystalPic}) make the fluid cloudier than necessary while providing little to no directed light scattering.} It can be useful (though not necessary for most applications) to \DB{narrow the distribution of particle sizes by filtering the fluid. One useful example of this, is the removal of the small solids by running the fluid through an unperforated coffee filter. The rheoscopic fluid is then reconstituted by re-suspending the larger platelets caught in the filter in water. This process significantly reduces the turbidity of the fluid, while maintaining its pearslescence, and dilutes the other surfactants in the shaving cream, making the fluid less prone to foaming. Larger clumps with sizes on the order of a few millimeters that form on the coffee filter can be removed by running the fluid through a fine mesh kitchen strainer.}

\begin{figure}
\def\svgwidth{3.37in}
\begingroup%
  \makeatletter%
  \providecommand\color[2][]{%
    \errmessage{(Inkscape) Color is used for the text in Inkscape, but the package 'color.sty' is not loaded}%
    \renewcommand\color[2][]{}%
  }%
  \providecommand\transparent[1]{%
    \errmessage{(Inkscape) Transparency is used (non-zero) for the text in Inkscape, but the package 'transparent.sty' is not loaded}%
    \renewcommand\transparent[1]{}%
  }%
  \providecommand\rotatebox[2]{#2}%
  \ifx\svgwidth\undefined%
    \setlength{\unitlength}{241.70742416bp}%
    \ifx\svgscale\undefined%
      \relax%
    \else%
      \setlength{\unitlength}{\unitlength * \real{\svgscale}}%
    \fi%
  \else%
    \setlength{\unitlength}{\svgwidth}%
  \fi%
  \global\let\svgwidth\undefined%
  \global\let\svgscale\undefined%
  \makeatother%
  \begin{picture}(1,1.18900956)%
    \put(0,0){\includegraphics[width=\unitlength,page=1]{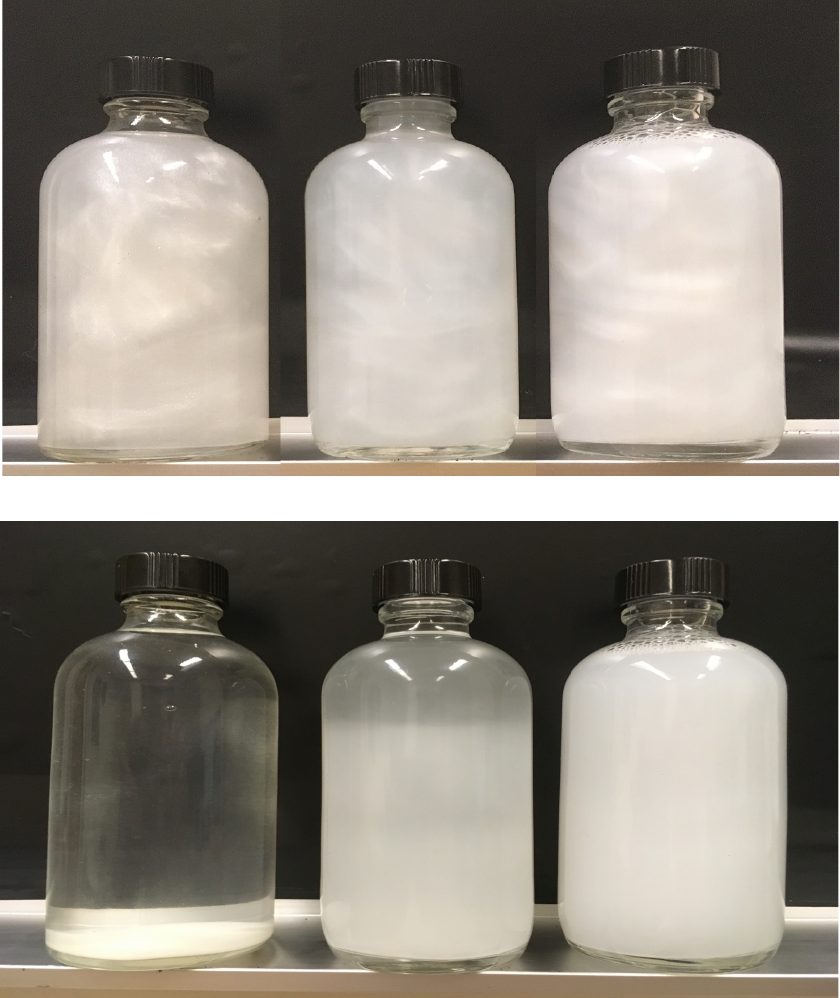}}%
    \put(0.03770595,1.1265632){\color[rgb]{1,1,1}\makebox(0,0)[lb]{\smash{(a)}}}%
    \put(0.03770595,0.49846493){\color[rgb]{1,1,1}\makebox(0,0)[lb]{\smash{(b)}}}%
  \end{picture}%
\endgroup%
\caption{\label{fig:settling}\DB{(a) Suspensions of mica flakes (left), \kscope\ (center), and stearic acid crystals (right) were thoroughly mixed. (b) After 12 hours, sedimentation is obvious for the mica and \kscope\ samples, while the stearic acid crystals remain in suspension.}}
\end{figure}

The main advantage of stearic acid platelets over other particles is that the density of stearic acid closely matches that of water. This allows stearic acid flakes to stay suspended much longer than mica or aluminum flakes, which have densities of 2.6-3.2 g/cm$^3$ and 2.7 g/cm$^3$, respectively.\cite{Haynes2017} \DB{Although the literature\cite{Biltz1930,Adriaanse1964} lists the density of stearic acid as 1.026 g/cm$^3$, in our experience stearic acid crystals extracted from shaving cream have a distribution of densities. Some crystals eventually settle to the bottom of the container, while others float to the surface of the fluid. Most likely this is because commercial stearic acid contains varying amounts of the less dense palmitic acid.\cite{CIR1987}} However, the sedimentation process takes weeks, vastly outperforming even \kscope\ in this regard, \DB{as can be seen in Fig. \ref{fig:settling}.}

\DB{The close density match between stearic acid and water also allows our fluid to react quickly to changes in flow conditions. Specifying the particle response time for the stearic acid crystals precisely is difficult, given their wide range of sizes and irregular shapes, but we can get an order of magnitude estimate by calculating the particle relaxation time for spherical particles of similar size. Using the longest dimension of the largest particles ($\sim 50\,\mu \mathrm{m}$) as the particle ``diameter'', we find that their particle relaxation time\cite{Raffel2018} 
\begin{equation}
\label{eq:responsetime}
\tau_0 = \left(\rho_p-\rho_f\right)\frac{d_p^2}{18 \mu_f}
\end{equation}
is on the order of 4 microseconds. Here, $\rho_p$ is the density of the particles, $d_p$ is their typical diameter, and $\rho_f$ and $\mu_f$ are the density and dynamic viscosity of the surrounding fluid, respectively. Since the particle relaxation time is proportional to the density mismatch between the particles and the fluid, $\tau_0$ is about 25 times shorter for stearic acid crystals than for similarly sized, but much denser, \kscope\ flakes. Naturally, since the particle relaxation time is proportional to the square of the particle diameter, even faster response times can be obtained by filtering out the largest particles.}

\DB{Stearic acid-based rheoscopic fluids have viscosities and densities very similar to those of water.} Using a Cannon-Fenske routine viscometer, we determined that the concentrated formulation discussed above has a viscosity that is only $\sim$5.5\% greater than that of water at 23$^\circ$C, meaning that the more dilute suspensions that would be used in experiments will have viscosities very close to those of water. Measurements of the concentrate's specific gravity using a hydrometer \DB{were indistinguishable from those for water to within the resolution of the instrument ($\pm 0.001$). This might be expected given the density of stearic acid and the relatively low volume fraction of the crystals within the suspension.}

Stearic acid is also useful in formulating rheoscopic fluids for the study of stratified flows, where it is common to use concentrated solutions of various salts. The salt ions can destroy the charged double layer that prevents \kscope\ particles from agglomerating, causing them to form spherical clumps and leading to a loss of the rheoscopic effect.\cite{Matisse1984} Stearic acid, being a waxy saturated fatty acid, resists this to a greater degree. While clumping still occurs for sufficiently high salt concentrations, this can be prevented by adding small amount of laureth-23 or other surfactant with a high hydrophilic-lypophilic balance (HLB). Using this method, we were able to formulate useful rheoscopic fluids with high concentrations of various salts, including NaCl, CuSO$_4$, and NH$_{4}$SCN.  

\DB{Rheoscopic fluids based on stearic acid flakes have significant advantages, but also have a few drawbacks.} Their main disadvantage stems from the fact that the melting point of stearic acid is relatively low\cite{Ralston1942} (69.2$^\circ$C) compared to that of guanine (360$^\circ$C) or mica ($>700^\circ$C).\cite{Haynes2017} This limits its use for higher temperature applications such as convection studies. Our experiments show that when stearic acid-based rheoscopic fluid is heated above 42$^\circ$C, it starts losing its pearlescence, losing it completely above $\sim$50$^\circ$C. We speculate that this may have to do with conformational changes between the various crystal phases of stearic acid that have been shown to occur in this temperature range prior to complete melting.\cite{Sato1987} \DB{Since the crystals must undergo a ripening process before they acquire the shape of platelets,\cite{Zhu2007} the fluid does not immediately regain the same level of pearlescence upon cooling and only becomes pearlescent again after approximately 24 hours.}

Another aspect in which stearic acid flakes are inferior to guanine and mica is that their index of refraction is only 1.43.\cite{Haynes2017} This makes them less effective scatterers, \DB{leading to slightly lower contrast images, as can be seen in Fig. \ref{fig:KvsB}, and limiting} their effectiveness in applications that exploit streaming birefringence, such as those discussed in Ref. \onlinecite{Hu2009}.

\section{Summary}
\label{sec:Summary}
We have demonstrated that stearic acid crystals extracted from common shaving cream can be used to formulate an excellent alternative to \kscope. Because stearic acid has a density almost identical to water, the sedimentation rate of the crystalline particles is very low and the fluid can maintain its rheoscopic properties almost indefinitely. \DB{The close density match also gives stearic acid-based rheoscopic fluids faster response times than other commonly used alternatives.} Stearic acid-based rheoscopic fluids can also be formulated for use in solutions of ionic salts, where other materials tend to clump and sediment or lose their rheoscopic properties. The low melting point of stearic acid limits the temperature range over which these fluids are useful, but for typical uses, they provide a versatile and inexpensive flow visualization tool.

\begin{acknowledgments}
D.B-E gratefully acknowledges the support of the M.J. Murdock Charitable Trust (Award \# 2015214) and the Kresge Endowment at Wi\-lla\-mette University. C.J.C. would like to acknowledge support from the National Science Foundation (Awards \# DMS-1125302 and CMMI-1234436). \DB{The authors appreciate the help of D. Altman, E.F. Greco, and L. Kageorge in acquiring the photographs in Figs. \ref{fig:CrystalPic}, \ref{fig:FlowViz}, and \ref{fig:settling}. We would also like to thank one of the referees, who brought up the issue of response time. Without their comment, the particularly fast reaction time of stearic acid crystals in water would have gone unnoticed by us.}
\end{acknowledgments}


\begin{thebibliography}{38}%
\makeatletter
\providecommand \@ifxundefined [1]{%
 \@ifx{#1\undefined}
}%
\providecommand \@ifnum [1]{%
 \ifnum #1\expandafter \@firstoftwo
 \else \expandafter \@secondoftwo
 \fi
}%
\providecommand \@ifx [1]{%
 \ifx #1\expandafter \@firstoftwo
 \else \expandafter \@secondoftwo
 \fi
}%
\providecommand \natexlab [1]{#1}%
\providecommand \enquote  [1]{``#1''}%
\providecommand \bibnamefont  [1]{#1}%
\providecommand \bibfnamefont [1]{#1}%
\providecommand \citenamefont [1]{#1}%
\providecommand \href@noop [0]{\@secondoftwo}%
\providecommand \href [0]{\begingroup \@sanitize@url \@href}%
\providecommand \@href[1]{\@@startlink{#1}\@@href}%
\providecommand \@@href[1]{\endgroup#1\@@endlink}%
\providecommand \@sanitize@url [0]{\catcode `\\12\catcode `\$12\catcode
  `\&12\catcode `\#12\catcode `\^12\catcode `\_12\catcode `\%12\relax}%
\providecommand \@@startlink[1]{}%
\providecommand \@@endlink[0]{}%
\providecommand \url  [0]{\begingroup\@sanitize@url \@url }%
\providecommand \@url [1]{\endgroup\@href {#1}{\urlprefix }}%
\providecommand \urlprefix  [0]{URL }%
\providecommand \Eprint [0]{\href }%
\providecommand \doibase [0]{http://dx.doi.org/}%
\providecommand \selectlanguage [0]{\@gobble}%
\providecommand \bibinfo  [0]{\@secondoftwo}%
\providecommand \bibfield  [0]{\@secondoftwo}%
\providecommand \translation [1]{[#1]}%
\providecommand \BibitemOpen [0]{}%
\providecommand \bibitemStop [0]{}%
\providecommand \bibitemNoStop [0]{.\EOS\space}%
\providecommand \EOS [0]{\spacefactor3000\relax}%
\providecommand \BibitemShut  [1]{\csname bibitem#1\endcsname}%
\let\auto@bib@innerbib\@empty
\bibitem [{\citenamefont {Westerweel}, \citenamefont {Elsinga},\ and\
  \citenamefont {Adrian}(2013)}]{Westerweel2013}%
  \BibitemOpen
  \bibfield  {author} {\bibinfo {author} {\bibfnamefont {J.}~\bibnamefont
  {Westerweel}}, \bibinfo {author} {\bibfnamefont {G.~E.}\ \bibnamefont
  {Elsinga}}, \ and\ \bibinfo {author} {\bibfnamefont {R.~J.}\ \bibnamefont
  {Adrian}},\ }\bibfield  {title} {\enquote {\bibinfo {title} {Particle image
  velocimetry for complex and turbulent flows},}\ }\href@noop {} {\bibfield
  {journal} {\bibinfo  {journal} {Annu. Rev. Fluid Mech.}\ }\textbf {\bibinfo
  {volume} {45}},\ \bibinfo {pages} {409--436} (\bibinfo {year}
  {2013})}\BibitemShut {NoStop}%
\bibitem [{\citenamefont {Raffel}\ \emph {et~al.}(2018)\citenamefont {Raffel},
  \citenamefont {Willert}, \citenamefont {Scarano}, \citenamefont {K{\"a}hler},
  \citenamefont {Wereley},\ and\ \citenamefont {Kompenhans}}]{Raffel2018}%
  \BibitemOpen
	\DB{
  \bibfield  {author} {\bibinfo {author} {\bibfnamefont {M.}~\bibnamefont
  {Raffel}}, \bibinfo {author} {\bibfnamefont {C.~E.}\ \bibnamefont {Willert}},
  \bibinfo {author} {\bibfnamefont {F.}~\bibnamefont {Scarano}}, \bibinfo
  {author} {\bibfnamefont {C.~J.}\ \bibnamefont {K{\"a}hler}}, \bibinfo
  {author} {\bibfnamefont {S.~T.}\ \bibnamefont {Wereley}}, \ and\ \bibinfo
  {author} {\bibfnamefont {J.}~\bibnamefont {Kompenhans}},\ }\href@noop {}
  {\emph {\bibinfo {title} {Particle image velocimetry: {A} practical
  guide}}},\ \bibinfo {edition} {2nd}\ ed.\ (\bibinfo  {publisher} {Springer
  International Publishing},\ \bibinfo {address} {New York, NY},\ \bibinfo
  {year} {2018})}\BibitemShut {NoStop}%
\bibitem [{\citenamefont {Merzkirch}(1987)}]{Merzkrich1987}%
  \BibitemOpen
  \bibfield  {author} {\bibinfo {author} {\bibfnamefont {W.}~\bibnamefont
  {Merzkirch}},\ }\href@noop {} {\emph {\bibinfo {title} {Flow
  Visualization}}},\ \bibinfo {edition} {2nd}\ ed.\ (\bibinfo  {publisher}
  {Academic Press Inc.},\ \bibinfo {address} {Cambridge},\ \bibinfo {year}
  {1987})\BibitemShut {NoStop}%
\bibitem [{\citenamefont {Smits}\ and\ \citenamefont {Lim}(2012)}]{Smits2012}%
  \BibitemOpen
  \bibinfo {editor} {\bibfnamefont {A.~J.}\ \bibnamefont {Smits}}\ and\
  \bibinfo {editor} {\bibfnamefont {T.~T.}\ \bibnamefont {Lim}},\ eds.,\
  \href@noop {} {\emph {\bibinfo {title} {Flow Visualization: Techniques and
  Examples}}},\ \bibinfo {edition} {2nd}\ ed.\ (\bibinfo  {publisher} {Imperial
  College Press},\ \bibinfo {address} {London},\ \bibinfo {year}
  {2012})\BibitemShut {NoStop}%
\bibitem [{\citenamefont {Weidman}(1989)}]{Weidman1989}%
  \BibitemOpen
  \bibfield  {author} {\bibinfo {author} {\bibfnamefont {P.~D.}\ \bibnamefont
  {Weidman}},\ }\href@noop {} {\emph {\bibinfo {title}
  {\textnormal{``Measurement Techniques in Laboratory Rotating Flows'' in}
  {A}dvances in {F}luid {M}echanics {M}easurements}}}\ (\bibinfo  {publisher}
  {Springer-Verlag},\ \bibinfo {address} {Heidelberg},\ \bibinfo {year}
  {1989})\ pp.\ \bibinfo {pages} {412--418}\BibitemShut {NoStop}%
\bibitem [{\citenamefont {Jeffery}(1922)}]{Jeffery1922}%
  \BibitemOpen\DB{
  \bibfield  {author} {\bibinfo {author} {\bibfnamefont {G.~B.}\ \bibnamefont
  {Jeffery}},\ }\bibfield  {title} {\enquote {\bibinfo {title} {The motion of
  ellipsoidal particles immersed in a viscous fluid},}\ }\href@noop {}
  {\bibfield  {journal} {\bibinfo  {journal} {Proc. R. Soc. Lond. A}\ }\textbf
  {\bibinfo {volume} {102}},\ \bibinfo {pages} {161--179} (\bibinfo {year}
  {1922})}}\BibitemShut {NoStop}%
\bibitem [{\citenamefont {Sava\c{s}}(1985)}]{Savas1985}%
  \BibitemOpen
  \bibfield  {author} {\bibinfo {author} {\bibfnamefont {{\"O}.}~\bibnamefont
  {Sava\c{s}}},\ }\bibfield  {title} {\enquote {\bibinfo {title} {On flow
  visualization using reflective flakes},}\ }\href@noop {} {\bibfield
  {journal} {\bibinfo  {journal} {J. Fluid Mech.}\ }\textbf {\bibinfo {volume}
  {152}},\ \bibinfo {pages} {235--248} (\bibinfo {year} {1985})}\BibitemShut
  {NoStop}%
\bibitem [{\citenamefont {Gauthier}, \citenamefont {Gondret},\ and\
  \citenamefont {Rabaud}(1998)}]{Gauthier1998}%
  \BibitemOpen
  \bibfield  {author} {\bibinfo {author} {\bibfnamefont {G.}~\bibnamefont
  {Gauthier}}, \bibinfo {author} {\bibfnamefont {P.}~\bibnamefont {Gondret}}, \
  and\ \bibinfo {author} {\bibfnamefont {M.}~\bibnamefont {Rabaud}},\
  }\bibfield  {title} {\enquote {\bibinfo {title} {Motions of anisotropic
  particles: Application to visualization of three-dimensional flows},}\
  }\href@noop {} {\bibfield  {journal} {\bibinfo  {journal} {Phys. Fluids}\
  }\textbf {\bibinfo {volume} {10}},\ \bibinfo {pages} {2147--2154} (\bibinfo
  {year} {1998})}\BibitemShut {NoStop}%
\bibitem [{\citenamefont {Goto}, \citenamefont {Kida},\ and\ \citenamefont
  {Fujiwara}(2011)}]{Goto2011}%
  \BibitemOpen
  \bibfield  {author} {\bibinfo {author} {\bibfnamefont {S.}~\bibnamefont
  {Goto}}, \bibinfo {author} {\bibfnamefont {S.}~\bibnamefont {Kida}}, \ and\
  \bibinfo {author} {\bibfnamefont {S.}~\bibnamefont {Fujiwara}},\ }\bibfield
  {title} {\enquote {\bibinfo {title} {Flow visualization using reflective
  flakes},}\ }\href@noop {} {\bibfield  {journal} {\bibinfo  {journal} {J.
  Fluid Mech.}\ }\textbf {\bibinfo {volume} {683}},\ \bibinfo {pages}
  {417--429} (\bibinfo {year} {2011})}\BibitemShut {NoStop}%
\bibitem [{\citenamefont {Shi}\ and\ \citenamefont {Mucha}(2014)}]{Shi2014}%
  \BibitemOpen\DB{
  \bibfield  {author} {\bibinfo {author} {\bibfnamefont {F.}~\bibnamefont
  {Shi}}\ and\ \bibinfo {author} {\bibfnamefont {P.~J.}\ \bibnamefont
  {Mucha}},\ }\bibfield  {title} {\enquote {\bibinfo {title} {Nonaxisymmetric
  high-aspect-ratio ellipsoids under shear: Lowest-order correction for finite
  aspect ratios},}\ }\href@noop {} {\bibfield  {journal} {\bibinfo  {journal}
  {Phys. Rev. E}\ }\textbf {\bibinfo {volume} {90}},\ \bibinfo {pages} {013005}
  (\bibinfo {year} {2014})}}\BibitemShut {NoStop}%
\bibitem [{\citenamefont {Byron}\ \emph {et~al.}(2015)\citenamefont {Byron},
  \citenamefont {Einarsson}, \citenamefont {Gustavsson}, \citenamefont {Voth},
  \citenamefont {Mehlig},\ and\ \citenamefont {Variano}}]{Byron2015}%
  \BibitemOpen\DB{
  \bibfield  {author} {\bibinfo {author} {\bibfnamefont {M.}~\bibnamefont
  {Byron}}, \bibinfo {author} {\bibfnamefont {J.}~\bibnamefont {Einarsson}},
  \bibinfo {author} {\bibfnamefont {K.}~\bibnamefont {Gustavsson}}, \bibinfo
  {author} {\bibfnamefont {G.}~\bibnamefont {Voth}}, \bibinfo {author}
  {\bibfnamefont {B.}~\bibnamefont {Mehlig}}, \ and\ \bibinfo {author}
  {\bibfnamefont {E.}~\bibnamefont {Variano}},\ }\bibfield  {title} {\enquote
  {\bibinfo {title} {Shape-dependence of particle rotation in isotropic
  turbulence},}\ }\href@noop {} {\bibfield  {journal} {\bibinfo  {journal}
  {Phys. Fluids}\ }\textbf {\bibinfo {volume} {27}},\ \bibinfo {pages} {035101}
  (\bibinfo {year} {2015})}}\BibitemShut {NoStop}%
\bibitem [{\citenamefont {Wesfreid}(2006)}]{Wesfreid2006a}%
  \BibitemOpen
  \bibfield  {author} {\bibinfo {author} {\bibfnamefont {J.~E.}\ \bibnamefont
  {Wesfreid}},\ }\href@noop {} {\emph {\bibinfo {title}
  {\textnormal{``Scientific Biography of Henri B\'{e}nard (1874\hspace{0.02
  in}--1939)'' in} Dynamics of Spatio-Temporal Cellular Structures}}}\
  (\bibinfo  {publisher} {Springer-Verlag},\ \bibinfo {address} {New York},\
  \bibinfo {year} {2006})\ pp.\ \bibinfo {pages} {9--37}\BibitemShut {NoStop}%
\bibitem [{\citenamefont {Raffel}\ \emph {et~al.}(2007)\citenamefont {Raffel},
  \citenamefont {Willert}, \citenamefont {Wereley},\ and\ \citenamefont
  {Kompenhans}}]{Raffel2007}%
  \BibitemOpen
  \bibfield  {author} {\bibinfo {author} {\bibfnamefont {M.}~\bibnamefont
  {Raffel}}, \bibinfo {author} {\bibfnamefont {C.~E.}\ \bibnamefont {Willert}},
  \bibinfo {author} {\bibfnamefont {S.~T.}\ \bibnamefont {Wereley}}, \ and\
  \bibinfo {author} {\bibfnamefont {J.}~\bibnamefont {Kompenhans}},\
  }\href@noop {} {\emph {\bibinfo {title} {Particle Image Velocimetry: {A}
  Practical Guide}}},\ \bibinfo {edition} {2nd}\ ed.\ (\bibinfo  {publisher}
  {Springer-Verlag},\ \bibinfo {address} {Heidelberg},\ \bibinfo {year}
  {2007})\BibitemShut {NoStop}%
\bibitem [{\citenamefont {Schultz-Grunow}\ and\ \citenamefont
  {Hein}(1956)}]{SchultzGrunow1956}%
  \BibitemOpen
  \bibfield  {author} {\bibinfo {author} {\bibfnamefont {F.}~\bibnamefont
  {Schultz-Grunow}}\ and\ \bibinfo {author} {\bibfnamefont {H.}~\bibnamefont
  {Hein}},\ }\bibfield  {title} {\enquote {\bibinfo {title} {Beitrag zur
  {Couettestr\"{o}mung}},}\ }\href@noop {} {\bibfield  {journal} {\bibinfo
  {journal} {Z. Flugwissensch.}\ }\textbf {\bibinfo {volume} {4}},\ \bibinfo
  {pages} {28--30} (\bibinfo {year} {1956})}\BibitemShut {NoStop}%
\bibitem [{\citenamefont {Coles}(1965)}]{Coles1965}%
  \BibitemOpen
  \bibfield  {author} {\bibinfo {author} {\bibfnamefont {D.}~\bibnamefont
  {Coles}},\ }\bibfield  {title} {\enquote {\bibinfo {title} {Transition in
  circular {Couette} flow},}\ }\href@noop {} {\bibfield  {journal} {\bibinfo
  {journal} {J. Fluid Mech.}\ }\textbf {\bibinfo {volume} {21}},\ \bibinfo
  {pages} {385--425} (\bibinfo {year} {1965})}\BibitemShut {NoStop}%
\bibitem [{\citenamefont {Matisse}(1969)}]{Matisse1969}%
  \BibitemOpen
  \bibfield  {author} {\bibinfo {author} {\bibfnamefont {P.~H.~R.}\
  \bibnamefont {Matisse}},\ }\href@noop {} {\emph {\bibinfo {title}
  {\textnormal{``Graphic Display,'' U.S. Patent 3\-464\-132}}}}\ (\bibinfo
  {year} {1969})\BibitemShut {NoStop}%
\bibitem [{\citenamefont {Krishnamurti}(1970)}]{Krishnamurti1970}%
  \BibitemOpen
  \bibfield  {author} {\bibinfo {author} {\bibfnamefont {R.}~\bibnamefont
  {Krishnamurti}},\ }\bibfield  {title} {\enquote {\bibinfo {title} {On the
  transition to turbulent convection. {Part} 1. {The} transition from two- to
  three-dimensional flow},}\ }\href@noop {} {\bibfield  {journal} {\bibinfo
  {journal} {J. Fluid Mech.}\ }\textbf {\bibinfo {volume} {42}},\ \bibinfo
  {pages} {295--307} (\bibinfo {year} {1970})}\BibitemShut {NoStop}%
\bibitem [{\citenamefont {Borrero-Echeverry}(2014)}]{Borrero2014Thesis}%
  \BibitemOpen
  \bibfield  {author} {\bibinfo {author} {\bibfnamefont {D.}~\bibnamefont
  {Borrero-Echeverry}},\ }\emph {\bibinfo {title} {Subcritical Transition to
  Turbulence in {Taylor-Couette} Flow}},\ \href@noop {} {Ph.D. thesis},\
  \bibinfo  {school} {School of Physics, Georgia Institute of Technology}
  (\bibinfo {year} {2014})\BibitemShut {NoStop}%
\bibitem [{\citenamefont {Pfaff}\ and\ \citenamefont
  {Reynders}(1999)}]{Pfaff1999}%
  \BibitemOpen
  \bibfield  {author} {\bibinfo {author} {\bibfnamefont {G.}~\bibnamefont
  {Pfaff}}\ and\ \bibinfo {author} {\bibfnamefont {P.}~\bibnamefont
  {Reynders}},\ }\bibfield  {title} {\enquote {\bibinfo {title}
  {Angle-dependent optical effects deriving from submicron structures of films
  and pigments},}\ }\href@noop {} {\bibfield  {journal} {\bibinfo  {journal}
  {Chem. Rev.}\ }\textbf {\bibinfo {volume} {99}},\ \bibinfo {pages}
  {1963--1982} (\bibinfo {year} {1999})}\BibitemShut {NoStop}%
\bibitem [{\citenamefont {Matisse}\ and\ \citenamefont
  {Gorman}(1984)}]{Matisse1984}%
  \BibitemOpen
  \bibfield  {author} {\bibinfo {author} {\bibfnamefont {P.}~\bibnamefont
  {Matisse}}\ and\ \bibinfo {author} {\bibfnamefont {M.}~\bibnamefont
  {Gorman}},\ }\bibfield  {title} {\enquote {\bibinfo {title} {Neutrally
  buoyant anisotropic particles for flow visualization},}\ }\href@noop {}
  {\bibfield  {journal} {\bibinfo  {journal} {Phys. Fluids}\ }\textbf {\bibinfo
  {volume} {27}},\ \bibinfo {pages} {759--760} (\bibinfo {year}
  {1984})}\BibitemShut {NoStop}%
\bibitem [{\citenamefont {Crombie}(1997)}]{Crombie1997}%
  \BibitemOpen
  \bibfield  {author} {\bibinfo {author} {\bibfnamefont {R.~L.}\ \bibnamefont
  {Crombie}},\ }\bibfield  {title} {\enquote {\bibinfo {title} {Cold pearl
  surfactant-based blends},}\ }\href@noop {} {\bibfield  {journal} {\bibinfo
  {journal} {Int. J. Cosmet. Sci.}\ }\textbf {\bibinfo {volume} {19}},\
  \bibinfo {pages} {205--214} (\bibinfo {year} {1997})}\BibitemShut {NoStop}%
\bibitem [{\citenamefont {{Kalliroscope Corporation}}(2017)}]{Kscope2017}%
  \BibitemOpen
  \bibfield  {author} {\bibinfo {author} {\bibnamefont {{Kalliroscope
  Corporation}}},\ }\href@noop {} {\bibfield  {journal} {\bibinfo  {journal}
  {(Private communication)}\ } (\bibinfo {year} {May 15, 2017})}\BibitemShut
  {NoStop}%
\bibitem [{\citenamefont {Dominguez-Lerma}, \citenamefont {Ahlers},\ and\
  \citenamefont {Cannell}(1985)}]{Dominguez-Lerma1985}%
  \BibitemOpen
  \bibfield  {author} {\bibinfo {author} {\bibfnamefont {M.~A.}\ \bibnamefont
  {Dominguez-Lerma}}, \bibinfo {author} {\bibfnamefont {G.}~\bibnamefont
  {Ahlers}}, \ and\ \bibinfo {author} {\bibfnamefont {D.~S.}\ \bibnamefont
  {Cannell}},\ }\bibfield  {title} {\enquote {\bibinfo {title} {Effects of
  `{Kalliroscope}' flow visualization particles on rotating {Couette}-{Taylor}
  flow},}\ }\href@noop {} {\bibfield  {journal} {\bibinfo  {journal} {Phys.
  Fluids}\ }\textbf {\bibinfo {volume} {28}},\ \bibinfo {pages} {1204--1206}
  (\bibinfo {year} {1985})}\BibitemShut {NoStop}%
\bibitem [{\citenamefont {Hu}, \citenamefont {Goreau},\ and\ \citenamefont
  {Bush}(2009)}]{Hu2009}%
  \BibitemOpen
  \bibfield  {author} {\bibinfo {author} {\bibfnamefont {D.~L.}\ \bibnamefont
  {Hu}}, \bibinfo {author} {\bibfnamefont {T.~J.}\ \bibnamefont {Goreau}}, \
  and\ \bibinfo {author} {\bibfnamefont {J.~W.~M.}\ \bibnamefont {Bush}},\
  }\bibfield  {title} {\enquote {\bibinfo {title} {Flow visualization using
  tobacco mosaic virus},}\ }\href@noop {} {\bibfield  {journal} {\bibinfo
  {journal} {Exp. Fluids}\ }\textbf {\bibinfo {volume} {46}},\ \bibinfo {pages}
  {477--484} (\bibinfo {year} {2009})}\BibitemShut {NoStop}%
\bibitem [{\citenamefont {Marcos}\ \emph {et~al.}(2011)\citenamefont {Marcos},
  \citenamefont {Seymour}, \citenamefont {Luhar}, \citenamefont {Durham},
  \citenamefont {Mitchell}, \citenamefont {Macke},\ and\ \citenamefont
  {Stocker}}]{Marcos2011}%
  \BibitemOpen
  \bibfield  {author} {\bibinfo {author} {\bibnamefont {Marcos}}, \bibinfo
  {author} {\bibfnamefont {J.~S.}\ \bibnamefont {Seymour}}, \bibinfo {author}
  {\bibfnamefont {M.}~\bibnamefont {Luhar}}, \bibinfo {author} {\bibfnamefont
  {W.~M.}\ \bibnamefont {Durham}}, \bibinfo {author} {\bibfnamefont {J.~G.}\
  \bibnamefont {Mitchell}}, \bibinfo {author} {\bibfnamefont {A.}~\bibnamefont
  {Macke}}, \ and\ \bibinfo {author} {\bibfnamefont {R.}~\bibnamefont
  {Stocker}},\ }\bibfield  {title} {\enquote {\bibinfo {title} {Microbial
  alignment in flow changes ocean light climate},}\ }\href@noop {} {\bibfield
  {journal} {\bibinfo  {journal} {Proc. Nat. Acad. Sci. U.S.A.}\ }\textbf
  {\bibinfo {volume} {108}},\ \bibinfo {pages} {3860--3864} (\bibinfo {year}
  {2011})}\BibitemShut {NoStop}%
\bibitem [{\citenamefont {Abcha}\ \emph {et~al.}(2008)\citenamefont {Abcha},
  \citenamefont {Latrache}, \citenamefont {Crumeyrolle},\ and\ \citenamefont
  {Mutabazi}}]{Abcha2008}%
  \BibitemOpen
  \bibfield  {author} {\bibinfo {author} {\bibfnamefont {N.}~\bibnamefont
  {Abcha}}, \bibinfo {author} {\bibfnamefont {N.}~\bibnamefont {Latrache}},
  \bibinfo {author} {\bibfnamefont {O.}~\bibnamefont {Crumeyrolle}}, \ and\
  \bibinfo {author} {\bibfnamefont {I.}~\bibnamefont {Mutabazi}},\ }\bibfield
  {title} {\enquote {\bibinfo {title} {Qualitative relation between reflected
  light intensity by {Kalliroscope} flakes and velocity field in the
  {Couette-Taylor} flow system},}\ }\href@noop {} {\bibfield  {journal}
  {\bibinfo  {journal} {Exp. Fluids}\ }\textbf {\bibinfo {volume} {45}},\
  \bibinfo {pages} {85--94} (\bibinfo {year} {2008})}\BibitemShut {NoStop}%
\bibitem [{\citenamefont {Barth}\ and\ \citenamefont
  {Burns}(2007)}]{Barth2007}%
  \BibitemOpen
  \bibfield  {author} {\bibinfo {author} {\bibfnamefont {W.~L.}\ \bibnamefont
  {Barth}}\ and\ \bibinfo {author} {\bibfnamefont {C.~A.}\ \bibnamefont
  {Burns}},\ }\bibfield  {title} {\enquote {\bibinfo {title} {Virtual
  rheoscopic fluids for flow visualization},}\ }\href@noop {} {\bibfield
  {journal} {\bibinfo  {journal} {IEEE Trans. Vis. Comput. Graph.}\ }\textbf
  {\bibinfo {volume} {13}},\ \bibinfo {pages} {1751--1758} (\bibinfo {year}
  {2007})}\BibitemShut {NoStop}%
\bibitem [{\citenamefont {Hecht}, \citenamefont {Mucha},\ and\ \citenamefont
  {Turk}(2010)}]{Hecht2010}%
  \BibitemOpen
  \bibfield  {author} {\bibinfo {author} {\bibfnamefont {F.}~\bibnamefont
  {Hecht}}, \bibinfo {author} {\bibfnamefont {P.~J.}\ \bibnamefont {Mucha}}, \
  and\ \bibinfo {author} {\bibfnamefont {G.}~\bibnamefont {Turk}},\ }\bibfield
  {title} {\enquote {\bibinfo {title} {Virtual rheoscopic fluids},}\
  }\href@noop {} {\bibfield  {journal} {\bibinfo  {journal} {IEEE Trans. Vis.
  Comput. Graph.}\ }\textbf {\bibinfo {volume} {16}},\ \bibinfo {pages}
  {147--160} (\bibinfo {year} {2010})}\BibitemShut {NoStop}%
\bibitem [{\citenamefont {Zhu}\ \emph {et~al.}(2007)\citenamefont {Zhu},
  \citenamefont {Pudney}, \citenamefont {Heppenstall-Butler}, \citenamefont
  {Butler}, \citenamefont {Ferdinando},\ and\ \citenamefont
  {Kirkland}}]{Zhu2007}%
  \BibitemOpen
  \bibfield  {author} {\bibinfo {author} {\bibfnamefont {S.}~\bibnamefont
  {Zhu}}, \bibinfo {author} {\bibfnamefont {P.~D.~A.}\ \bibnamefont {Pudney}},
  \bibinfo {author} {\bibfnamefont {M.}~\bibnamefont {Heppenstall-Butler}},
  \bibinfo {author} {\bibfnamefont {M.~F.}\ \bibnamefont {Butler}}, \bibinfo
  {author} {\bibfnamefont {D.}~\bibnamefont {Ferdinando}}, \ and\ \bibinfo
  {author} {\bibfnamefont {M.}~\bibnamefont {Kirkland}},\ }\bibfield  {title}
  {\enquote {\bibinfo {title} {Interaction of the acid soap of triethanolamine
  stearate and stearic acid with water},}\ }\href@noop {} {\bibfield  {journal}
  {\bibinfo  {journal} {J. Phys. Chem. B}\ }\textbf {\bibinfo {volume} {111}},\
  \bibinfo {pages} {1016--1024} (\bibinfo {year} {2007})}\BibitemShut {NoStop}%
\bibitem [{\citenamefont {{K. Krishan, Procter \& Gamble
  Co.}}(2018)}]{Krishan2017}%
  \BibitemOpen\DB{
  \bibfield  {author} {\bibinfo {author} {\bibnamefont {{K. Krishan, Procter \&
  Gamble Co.}}},\ }\href@noop {} {\bibfield  {journal} {\bibinfo  {journal}
  {(Private communication)}\ } (\bibinfo {year} {July 28, 2018})}}\BibitemShut
  {NoStop}%
\bibitem [{EAF(2009)}]{EAF2009}%
  \BibitemOpen\DB{
  \href@noop {} {\enquote {\bibinfo {title} {Guide on hot water bath testing
  and its alternatives},}\ }\bibinfo {type} {Tech. Rep.}\ (\bibinfo
  {institution} {European Aerosol Federation},\ \bibinfo {address} {Avenue
  Herrmann-Debroux, 15A - 1160 Brussels, Belgium},\ \bibinfo {year}
  {2009})}\BibitemShut {NoStop}%
\bibitem [{\citenamefont {Sato}(1987)}]{Sato1987}%
  \BibitemOpen
  \bibfield  {author} {\bibinfo {author} {\bibfnamefont {K.}~\bibnamefont
  {Sato}},\ }\bibfield  {title} {\enquote {\bibinfo {title} {Physical and
  molecular properties of lipid polymorphs - {A} review},}\ }\href@noop {}
  {\bibfield  {journal} {\bibinfo  {journal} {Food Microstruct.}\ }\textbf
  {\bibinfo {volume} {6}},\ \bibinfo {pages} {151--159} (\bibinfo {year}
  {1987})}\BibitemShut {NoStop}%
\bibitem [{Note1()}]{Note1}%
  \BibitemOpen\DB{
  \bibinfo {note} {\protect This is approximately 1:1 by volume
  but measuring the volume of shaving cream is difficult, so we provide a more
  precise formulation in terms mass.}}\BibitemShut {Stop}%
\bibitem [{\citenamefont {Haynes}(2017)}]{Haynes2017}%
  \BibitemOpen
  \bibinfo {editor} {\bibfnamefont {W.~M.}\ \bibnamefont {Haynes}},\ ed.,\
  \href@noop {} {\emph {\bibinfo {title} {CRC Handbook of Chemistry and
  Physics}}},\ \bibinfo {edition} {97th}\ ed.\ (\bibinfo  {publisher} {CRC
  Press/Taylor \& Francis},\ \bibinfo {address} {Boca Raton},\ \bibinfo {year}
  {2017})\BibitemShut {NoStop}%
\bibitem [{\citenamefont {Biltz}, \citenamefont {Fischer},\ and\ \citenamefont
  {W{\"u}nnenberg}(1930)}]{Biltz1930}%
  \BibitemOpen
  \bibfield  {author} {\bibinfo {author} {\bibfnamefont {W.}~\bibnamefont
  {Biltz}}, \bibinfo {author} {\bibfnamefont {W.}~\bibnamefont {Fischer}}, \
  and\ \bibinfo {author} {\bibfnamefont {E.}~\bibnamefont {W{\"u}nnenberg}},\
  }\bibfield  {title} {\enquote {\bibinfo {title} {{\"U}ber {M}olekular- und
  {A}tomvolumina - 25. {{\"U}ber} die {R}aumbeanspruchung kristallisierter
  organischer {S}toffe bei tiefen {T}emperaturen},}\ }\href@noop {} {\bibfield
  {journal} {\bibinfo  {journal} {Z. Phys. Chem.}\ }\textbf {\bibinfo {volume}
  {151A}},\ \bibinfo {pages} {13--55} (\bibinfo {year} {1930})}\BibitemShut
  {NoStop}%
\bibitem [{\citenamefont {Adriaanse}, \citenamefont {Dekker},\ and\
  \citenamefont {Coops}(1964)}]{Adriaanse1964}%
  \BibitemOpen
  \bibfield  {author} {\bibinfo {author} {\bibfnamefont {N.}~\bibnamefont
  {Adriaanse}}, \bibinfo {author} {\bibfnamefont {H.}~\bibnamefont {Dekker}}, \
  and\ \bibinfo {author} {\bibfnamefont {J.}~\bibnamefont {Coops}},\ }\bibfield
   {title} {\enquote {\bibinfo {title} {Some physical constants of normal,
  saturated fatty acids and their methyl esters},}\ }\href@noop {} {\bibfield
  {journal} {\bibinfo  {journal} {Recl. Trav. Chim. Pays-Bas}\ }\textbf
  {\bibinfo {volume} {83}},\ \bibinfo {pages} {557--572} (\bibinfo {year}
  {1964})}\BibitemShut {NoStop}%
\bibitem [{\citenamefont {{Cosmetic Ingredient Review}}(1987)}]{CIR1987}%
  \BibitemOpen\DB{
  \bibfield  {author} {\bibinfo {author} {\bibnamefont {{Cosmetic Ingredient
  Review}}},\ }\bibfield  {title} {\enquote {\bibinfo {title} {Final report on
  the safety assessment of oleic acid, lauric acid, palmitic acid, myristic
  acid, and stearic acid},}\ }\href@noop {} {\bibfield  {journal} {\bibinfo
  {journal} {J. Am. Coll. Toxicol.}\ }\textbf {\bibinfo {volume} {6}},\
  \bibinfo {pages} {321--401} (\bibinfo {year} {1987})}}\BibitemShut {NoStop}%
\bibitem [{\citenamefont {Ralston}\ and\ \citenamefont
  {Hoerr}(1942)}]{Ralston1942}%
  \BibitemOpen
  \bibfield  {author} {\bibinfo {author} {\bibfnamefont {A.~W.}\ \bibnamefont
  {Ralston}}\ and\ \bibinfo {author} {\bibfnamefont {C.~W.}\ \bibnamefont
  {Hoerr}},\ }\bibfield  {title} {\enquote {\bibinfo {title} {The solubilities
  of normal saturated fatty acids},}\ }\href@noop {} {\bibfield  {journal}
  {\bibinfo  {journal} {J. Org. Chem.}\ }\textbf {\bibinfo {volume} {7}},\
  \bibinfo {pages} {546--555} (\bibinfo {year} {1942})}\BibitemShut {NoStop}%
\end{thebibliography}
\end{document}